\documentstyle[12pt,twoside]{article}
 
\def\a{\alpha}
\def\b{\beta}

\def\d{\delta}
\def\e{\epsilon}                
\def\f{\phi}                    

\def\h{\eta}

\def\k{\kappa}
\def\l{\lambda}
\def\m{\mu}
\def\n{\nu}
\def\o{\omega}
\def\p{\pi}                     
\def\r{\rho}                    
\def\s{\sigma}                  
\def\t{\tau}

\def\x{\xi}

\def\D{\Delta}

\def\G{\Gamma}

\def\cf{{\cal F}}

\def\ci{{\cal I}}

\def\ct{{\cal T}}

\def\bo{{\raise.05ex\hbox{\large$\Box$}\:}}             
\def\cbo{{\,\raise-.15ex\Sc [\,}}                       
\def\pa{\partial}                                       
\def\su{\sum}                                           
\def\TH{{\raise.2ex\hbox{$\displaystyle \bigodot$}\mskip-4.7mu \llap H \;}}
\def\face{\hbox{\normalsize$\;\;\:{\raise.9ex\hbox{\oo n}\mskip-13mu \llap
        {${\buildrel{\hbox{\frtnrm ..}}\over\smile}$}}\:$}}     
\def\Face{{\raise.2ex\hbox{$\displaystyle \bigodot$}\mskip-2.2mu \llap {$\ddot
        \smile$}}}                                      
\def\Lhat{{\bf\rlap{\kern-.09em$\hat{\phantom L}$}L}}
\def\Lcheck{{\bf\rlap{\kern-.09em$\check{\phantom L}$}L}}
 
\def\sp#1{{}^{#1}}                              
\def\sbra#1{\left\langle #1\right|}             
\def\sket#1{\left| #1\right\rangle}             
\def\svev#1{\left\langle #1\right\rangle}       
\def\leftrightarrowfill{$\mathsurround=0pt \mathord\leftarrow \mkern-6mu
        \cleaders\hbox{$\mkern-2mu \mathord- \mkern-2mu$}\hfill
        \mkern-6mu \mathord\rightarrow$}
\def\dvec#1{\vbox{\ialign{##\crcr
        \leftrightarrowfill\crcr\noalign{\kern-1pt\nointerlineskip}
        $\hfil\displaystyle{#1}\hfil$\crcr}}}           
\def\ddt#1{{\buildrel {\hbox{\LARGE .\kern-2pt.}} \over {#1}}}
 
\def\frac#1#2{{\textstyle{#1\over\vphantom2\smash{\raise.20ex
        \hbox{$\scriptstyle{#2}$}}}}}                   
\def\ha{\frac12}                                        
\def\sfrac#1#2{{\vphantom1\smash{\lower.5ex\hbox{\small$#1$}}\over
        \vphantom1\smash{\raise.4ex\hbox{\small$#2$}}}} 
\def\bfrac#1#2{{\vphantom1\smash{\lower.5ex\hbox{$#1$}}\over
        \vphantom1\smash{\raise.3ex\hbox{$#2$}}}}       
\def\afrac#1#2{{\vphantom1\smash{\lower.5ex\hbox{$#1$}}\over#2}}    
 
\def\boxes#1{
        \newcount\num
        \num=1
        \newdimen\downsy
        \downsy=-1.64ex
        \mskip-7.8mu
        \bo
        \loop
        \ifnum\num<#1
        \llap{\raise\num\downsy\hbox{$\bo$}}
        \advance\num by1
        \repeat}
\def\boxup#1#2{\newcount\numup
        \numup=#1
        \advance\numup by-1
        \newdimen\upsy
        \upsy=.82ex
        \mskip7.8mu
        \raise\numup\upsy\hbox{$#2$}}
 
\newskip\humongous \humongous=0pt plus 1000pt minus 1000pt
\def\caja{\mathsurround=0pt}

\newif\ifdtup
\def\panorama{\global\dtuptrue \openup2\jot \caja
        \everycr{\noalign{\ifdtup \global\dtupfalse
        \vskip-\lineskiplimit \vskip\normallineskiplimit
        \else \penalty\interdisplaylinepenalty \fi}}}
\def\li#1{\panorama \tabskip=\humongous                         
        \halign to\displaywidth{\hfil$\displaystyle{##}$
        \tabskip=0pt&$\displaystyle{{}##}$\hfil
        \tabskip=\humongous&\llap{$##$}\tabskip=0pt
        \crcr#1\crcr}}

\def\CMP{Commun. Math. Phys.}
\def\NP{Nucl. Phys. B}
\def\PL{Phys. Lett. }

\def\PRD{Phys. Rev. D}
\def\CQG{Class. Quant. Grav.}

\def\ref#1{$\sp{#1]}$}

\parskip=\medskipamount                 
\def\baselinestretch{1.2}       
\thicklines                         
\def\title#1#2#3#4{
\begin{document}
        {\hbox to\hsize{#4 \hfill  #3}}\par
        \begin{center}\vskip.5in minus.1in {\Large\bf #1}\\[.5in minus.2in]{#2}
        \vskip1.4in minus1.2in {\bf ABSTRACT}\\[.1in]\end{center}
        \begin{quotation}\par}
\def\author#1#2{#1\\[.1in]{\it #2}\\[.1in]}

\def\AMIC{Aleksandar Mikovic\'c
\\[.1in]{\it Blackett Laboratory, Imperial College, Prince Consort Road, London
SW7 2BZ, UK}\\[.1in]}

\def\AMICIF{Aleksandar Mikovi\'c\,
\footnote{Work supported by MNTRS and Royal Society}
\\[.1in] {\it Blackett Laboratory, Imperial College, Prince Consort
Road, London SW7 2BZ, UK}\\[.1in]
and \\[.1 in]
{\it Institute of Physics, P.O. Box 57, 11001 Belgrade, Yugoslavia}
\footnote{Permanent address}\\ {\it E-mail:\, mikovic@castor.phy.bg.ac.yu}}

\def\AMSISSA{Aleksandar Mikovi\'c\,
\footnote{E-mail address: mikovic@castor.phy.bg.ac.yu}
\\[.1in] {\it SISSA-International School for Advanced Studies\\
Via Beirut 2-4, Trieste 34100, Italy}\\[.1in]
and \\[.1 in]
{\it Institute of Physics, P.O. Box 57, 11001 Belgrade, Yugoslavia}
\footnote{Permanent address}}

\def\AM{Aleksandar Mikovi\'c 
\footnote{E-mail address: mikovic@castor.phy.bg.ac.yu}
\\[.1in] {\it Institute of Physics, P.O.Box 57, Belgrade 11001, Yugoslavia}
\\[.1in]}

\def\AMsazda{Aleksandar Mikovi\'c 
\footnote{E-mail address: mikovic@castor.phy.bg.ac.yu}
and Branislav Sazdovi\'c \footnote{E-mail: sazdovic@castor.phy.bg.ac.yu}
\footnote{Work supported by MNTRS}
\\[.1in] {\it Institute of Physics, P.O.Box 57, Belgrade 11001, Yugoslavia}
\\[.1in]}

\def\AMVR{Aleksandar Mikovi\'c\,
\footnote{E-mail address: mikovic@castor.phy.bg.ac.yu}
\\[.1in] 
{\it Institute of Physics, P.O. Box 57, 11001 Belgrade, Yugoslavia}
\\[.2in]
Voja Radovanovi\'c \\[.1 in]
{\it Faculty of Physics, P.O. Box 550, 11001 Belgrade, Yugoslavia}}

\def\AMCVR{Aleksandar Mikovi\'c
\footnote{Permanent address: Institute of Physics, P.O. Box 57, 11001 
Belgrade, Yugoslavia}\footnote{E-mail: mikovic@fy.chalmers.se, 
mikovic@castor.phy.bg.ac.yu}
\\
{\it Institute of Theoretical Physics, Chalmers University of Technology,
S-412 96 Goteborg, Sweden}\\[.1in]
and
\\[.1in]
Voja Radovanovi\'c
\footnote{E-mail: rvoja@rudjer.ff.bg.ac.yu} \\
{\it Faculty of Physics, P.O. Box 550, 11001 Belgrade, Yugoslavia}}

\def\AMVVR{Aleksandar Mikovi\'c
\footnote{On leave from Institute of Physics, P.O. Box 57, 11001 
Belgrade, Yugoslavia}
\footnote{Supported by Comissi\'on Interministerial de Ciencia y Tecnologia}
\footnote{E-mail: mikovic@lie1.ific.uv.es}
\\
{\it Departamento de Fisica Te\'orica and IFIC, Centro Mixto Universidad
de Valencia-CSIC, Facultad de Fisica, Burjassot-46100, Valencia, Spain}
\\[.1in]
Voja Radovanovi\'c
\footnote{E-mail: rvoja@rudjer.ff.bg.ac.yu} \\
{\it Faculty of Physics, P.O. Box 368, 11001 Belgrade, Yugoslavia}}

\def\endtitle{\par\end{quotation}\vskip3.5in minus2.3in\newpage}
 
 
\def\endabstract{\par\end{quotation}
        \renewcommand{\baselinestretch}{1.2}\small\normalsize}
 
 
\def\xpar{\par}                                         

\def\letterhead{
        \centerline{\large\sf INSTITUTE OF PHYSICS}
        \centerline{\sf P.O.Box 57, 11001 Belgrade, Yugoslavia}
        \rightline{\scriptsize\sf Dr Aleksandar Mikovi\'c}
        \vskip-.07in
        \rightline{\scriptsize\sf Tel: 11 615 575}
        \vskip-.07in
        \rightline{\scriptsize\sf E-mail: MIKOVIC@CASTOR.PHY.BG.AC.YU}}

\def\sig#1{{\leftskip=3.75in\parindent=0in\goodbreak\bigskip{Sincerely yours,}
\nobreak\vskip .7in{#1}\par}}

\def\ssig#1{{\leftskip=3.75in\parindent=0in\goodbreak\bigskip{}
\nobreak\vskip .7in{#1}\par}}

 
\def\ree#1#2#3{
        \hfuzz=35pt\hsize=5.5in\textwidth=5.5in
        \begin{document}
        \ttraggedright
        \par
        \noindent Referee report on Manuscript \##1\\
        Title: #2\\
        Authors: #3}
 
 
\def\start#1{\pagestyle{myheadings}\begin{document}\thispagestyle{myheadings}
        \setcounter{page}{#1}}
 
 
\catcode`@=11
 
\def\ps@myheadings{\def\@oddhead{\hbox{}\footnotesize\bf\rightmark \hfil
        \thepage}\def\@oddfoot{}\def\@evenhead{\footnotesize\bf
        \thepage\hfil\leftmark\hbox{}}\def\@evenfoot{}
        \def\sectionmark##1{}\def\subsectionmark##1{}
        \topmargin=-.35in\headheight=.17in\headsep=.35in}
\def\ps@acidheadings{\def\@oddhead{\hbox{}\rightmark\hbox{}}
        \def\@oddfoot{\rm\hfil\thepage\hfil}
        \def\@evenhead{\hbox{}\leftmark\hbox{}}\let\@evenfoot\@oddfoot
        \def\sectionmark##1{}\def\subsectionmark##1{}
        \topmargin=-.35in\headheight=.17in\headsep=.35in}
 
\catcode`@=12
 
\def\sect#1{\bigskip\medskip\goodbreak\noindent{\large\bf{#1}}\par\nobreak
        \medskip\markright{#1}}
\def\chsc#1#2{\phantom m\vskip.5in\noindent{\LARGE\bf{#1}}\par\vskip.75in
        \noindent{\large\bf{#2}}\par\medskip\markboth{#1}{#2}}
\def\Chsc#1#2#3#4{\phantom m\vskip.5in\noindent\halign{\LARGE\bf##&
        \LARGE\bf##\hfil\cr{#1}&{#2}\cr\noalign{\vskip8pt}&{#3}\cr}\par\vskip
        .75in\noindent{\large\bf{#4}}\par\medskip\markboth{{#1}{#2}{#3}}{#4}}
\def\chap#1{\phantom m\vskip.5in\noindent{\LARGE\bf{#1}}\par\vskip.75in
        \markboth{#1}{#1}}
\def\refs{\bigskip\medskip\goodbreak\noindent{\large\bf{REFERENCES}}\par
        \nobreak\bigskip\markboth{REFERENCES}{REFERENCES}
        \frenchspacing \parskip=0pt \renewcommand{\baselinestretch}{1}\small}
\def\unrefs{\normalsize \nonfrenchspacing \parskip=medskipamount}
\def\Item{\par\hang\textindent}
\def\Itemitem{\par\indent \hangindent2\parindent \textindent}
\def\makelabel#1{\hfil #1}
\def\topic{\par\noindent \hangafter1 \hangindent20pt}
\def\Topic{\par\noindent \hangafter1 \hangindent60pt}

\title{Loop Corrections in the Spectrum of 2D Hawking Radiation} 
{\AMVVR}{FTUV/97-11, IFIC/97-11}{March 1997}

We determine the one-loop and the two-loop back-reaction corrections in the 
spectrum of the Hawking
radiation for the CGHS model of 2d dilaton gravity by evaluating the
Bogoliubov coefficients for a massless scalar field propagating on the
corresponding backgrounds. 
Since the back-reaction can induce a small shift in the position of the 
classical
horizon, we find that a positive shift leads to a non-Planckian late-time
spectrum, while a null or a negative shift leads to a Planckian late-time
spectrum in the leading-order stationary-point approximation. 
In the one-loop case there are no corrections to the classical
Hawking temperature, while in the two-loop case the temperature is
three times greater than the classical value.
We argue that these results are consistent with the behaviour of
the Hawking flux obtained from the operator quantization only for the times 
which are not too late, 
in accordance with the limits of validity of the semiclassical approximation. 

\endtitle

\sect{1. Introduction}

Two-dimensional (2d) dilaton gravity models have turned out to be excellent 
toy models for understanding black hole formation and back-reaction of the
Hawking radiation (for a review see \cite{rev}). The CGHS model \cite{cghs}
has been studied most extensively, due to its simplicity. In \cite{cghs}
it was shown that a thermal Hawking radiation can exist in the semiclassical 
limit, which was done by evaluating the Hawking flux from the trace anomaly. 
This result
was later confirmed by a direct calculation of the Bogoliubov coefficients
\cite{gn}, which was the 2d analog of the famous Hawking calculation 
\cite{hawk}. In a further development, explicit solutions 
which include the back-reaction were found, in one-loop
\cite{rst,m95,bpp} and two-loop \cite{mr} approximations.

In the one-loop case,
the trace anomaly method \cite{bpp}, as well as the point-splitting method
\cite{m96}, yield the same Hawking 
flux as in the zero-loop case, which is a good indication that the one-loop 
back-reaction does not change the Planckian spectrum of the radiation. 
This result is
surprising, because one expects that the back-reaction should change
the zero-loop temperature for a small amount. This is especially
puzzling in the case of the BPP solution \cite{bpp}, where the thermal 
Hawking flux gets turned off for late
times (although discontinuously), leaving behind a static remnant geometry.
Another puzzling feature of the BPP solution is that the corresponding
semiclassical geometry has a horizon which is shifted for a small
amount with respect to the classical horizon. The horizon shift also appears
in the two-loop case, while the operator quantization yields an 
expected result for the Hawking flux, 
in the sense that the zero-loop flux changes under the 
two-loop back-reaction \cite{mr}. 

These results make interesting the 
calculation of the Bogoliubov coefficients, since one would like to see
what kind of corrections these back-reaction effects would make in the 
spectrum of the Hawking radiation. Another interesting point to be examined
is that one can
evaluate the Hawking flux from the Bogoliubov coefficients and compare it
to the Hawking flux obtained from the expectation value of the energy-momentum
tensor operator. Therefore we are
going to evaluate the Bogoliubov coefficients for a free scalar field
propagating on the one-loop geometry of \cite{bpp}
and the two-loop geometry of \cite{mr}. 

In section two we briefly review the quantization of the CGHS model in the
operator formalism of \cite{m96}, and define the loop expansion. In section 
three we
evaluate the Bogoliubov coefficients for the BPP background geometry
and analyse the corresponding spectrum of the Hawking radiation.
The same is done in section five, but in the case of the two-loop background
geometry of ref. \cite{mr}. 
In section six we present our conclusions. In the appendix we
give some theorems which are useful for obtaining the asymptotic behaviour
of the Bogoliubov coefficients. 

\sect{2. Loop-expansion for 2d dilaton gravity}
  
Our starting point is a
classical theory described by the CGHS action \cite{cghs}
$$ S =  \int_{M} d^2 x \sqrt{-g} \left[ e^{-\f}\left( R + 
 (\nabla \f)^2 + 4\l^2 \right) - \ha (\nabla f)^2 \right]\quad,
\eqno(2.1)$$
where $\f$ is a dilaton scalar field, $f$ is a matter scalar field,  
$g$, $R$ and $\nabla$ are the 
determinant, the curvature scalar and the covariant derivative respectively,
associated with a metric $g_{\m\n}$ on a 2d manifold $M$. The topology of 
$M$ is that of $ {\bf R} \times {\bf R}$. The equations of motion can
be solved in the conformal gauge $ds^2 = -e^{\r}dx^+ dx^-$ \cite{cghs} as
$$e^{-\r} = e^{-\f} =  - \l^2 x^+ x^- - F_+ - F_- \quad,\quad
f = f_+ (x^+) + f_-(x^-) \quad, \eqno(2.2)$$
where
$$ F_{\pm}= a_{\pm} + b_{\pm}x^{\pm} + 
\int^{x^{\pm}} dy \int^y dz T_{\pm\pm} (z) 
\quad,\eqno(2.3)$$
and $T_{\pm\pm}$ is the matter energy-momentum tensor
$$T_{\pm\pm} = \ha \pa_{\pm} f \pa_{\pm} f \quad.\eqno(2.4)$$ 
The residual conformal invariance is fixed by a gauge choice $\r = \f$,
and the independent integration constants are $a_+ + a_-$ and $b_{\pm}$.

The quantum theory can be obtained by quantizing the space of classical
solutions defined by the equations (2.2) and (2.3) \cite{m96}. 
This is equivalent to a
reduced phase space quantization of the action (2.1) \cite{m95}. In this way 
one obtains a quantum theory of a
free massless scalar field $f$ propagating on a flat fictitious background 
$ds^2 = -dx^+ dx^-$. The dilaton and the conformal factor are
given by the expression in (2.2), which is considered as an operator 
in the Heisenberg picture. The matter energy-momentum tensor operator is
defined as
$$T_{\pm\pm} = \ha :\pa_{\pm} f \pa_{\pm} f: \quad,\eqno(2.5)$$
where the normal ordering in (2.5) is chosen to be with respect to
the vacuum for the creation
and annihilation operators $(a_k^{\dagger}, a_k )$ defined by
$$f_{\pm} (x^{\pm}) = {1\over\sqrt{2\p}}\int_{0}^{\infty}
{dk\over\sqrt{2\o_k}}\left[ a_{\mp k} e^{-ikx^{\pm}} + a_{\mp k}^{\dagger} 
e^{ikx^{\pm}}\right]\quad,\eqno(2.6)$$
where $\o_k = |k|$.

The physical Hilbert space of the model is 
the Fock space ${\cf} (a_k)$ constructed from $a_k^{\dagger}$ acting on
the vacuum $\sket{0}$. The model
is unitary because the dynamics is generated by a free-field
Hamiltonian 
$$H = \int_{-\infty}^{\infty} dk\, \o_k a_{k}^{\dagger} a_k + E_0 \quad,
\eqno(2.7)$$
which is a Hermitian operator acting on ${\cf}$, where
$E_0$ is the vacuum energy. Consequently the states at $t = \ha (x^+ + x^-)=$
const. surfaces are related by a unitary transformation
$$  \Psi (t_2) = e^{-iH(t_2 - t_1)}\Psi(t_1) \quad.\eqno(2.8)$$
The Heisenberg picture is defined as
$$ \Psi_0 = e^{iHt}\Psi(t) \quad,\quad A (t) = e^{iHt}Ae^{-iHt}\quad,
\eqno(2.9)$$
which also serves to relate the covariant quantization expressions
to the canonical quantization expressions. For example,
$$f(t,x) = e^{iHt} f(x) e^{-iHt} = {1\over\sqrt{2\p}}\int_{-\infty}^{\infty}
{dk\over\sqrt{2\o_k}}\left[ a_k e^{i(kx-\o_k t)} + a_k^{\dagger} 
e^{-i(kx-\o_k t)}\right]\quad,\eqno(2.10)$$
where $x = \ha (x^+ - x^-)$. 

Given a physical state $\Psi_0$, one can associate an effective metric to 
$\Psi (t)=e\sp{-iHt}\Psi_0$ via
$$e^{\r_{eff} (t,x)}=e^{\f_{eff} (t,x)}
=\sbra{\Psi (t)}e^{\f(x)}\sket{\Psi (t)} 
=\sbra{\Psi_0}e^{\f(t,x)}\sket{\Psi_0} \quad,\eqno(2.11)$$ 
where $e^{\r_{eff}}$ is the effective conformal factor.
The geometry which is generated by $e^{\r_{eff}}$ via 
$ds^2 = - e^{\r_{eff}}dx^+dx^-$ makes sense only in the regions of $M$
where the quantum fluctuations are small \cite{m95,m96}. The
condition for this is
$$ \sqrt{|\svev{e^{2\f}} - \svev{e^\f}^2 |} << \svev{e^\f}\quad,\eqno(2.12)$$
which defines the limits of validity of the semiclassical approximation.
At least perturbatively, the region defined by (2.12) roughly coincides
with the weak-coupling region \cite{m95,m96}.

The effective conformal factor can be calculated perturbatively
by using an expansion in powers of the energy-momentum tensor,
which is equivalent to the expansion in matter loops \cite{m95,m96}
$$\svev{(-\l^2 x^+ x^- -  F )^{-1}} = 
e^{\f_0} \su_{n=0}^{\infty} e^{n\f_0} \svev{\d F^n}\quad,\eqno(2.13)$$
where $F_0$ is a c-number function, $e^{-\f_0} = -\l^2 x^+ x^- -  F_0 $ and
$\d F = F - F_0$. 
A convenient choice for $F_0$ is
$$ F_0 = \svev{F_+} + \svev{F_-}\quad,\eqno(2.14) $$
since then the lowest order metric is a one-loop semiclassical metric
$$ e^{-\f_0} = -\l^2 x^+ x^- -  \svev{F_+} - \svev{F_-}\quad.\eqno(2.15) $$

The state $\Psi_0$ is chosen such that it is as close as possible to the
classical matter distribution $f_0 (x^+)$ which creates the black hole. 
The corresponding classical metric is described by
$$ e^{-\r} = {M(x^+)\over \l} - \l^2 x^+ \D (x^+) - \l^2 x^+ x^-  
\eqno(2.16)$$
where
$$ M (x^+) = \l\int_{-\infty}^{x^+} dy\, y \,T_{++}^0 (y)\quad,\quad
\l^2 \D = \int_{-\infty}^{x^+} dy\, T^0_{++} (y) \quad \eqno(2.17)$$
and $T_{++}^0 = \ha \pa_{+}f_0 \pa_{+} f_0$. The geometry is that of a black 
hole of the mass 
$$M = \lim_{x^+ \to +\infty} M(x^+)\quad,\eqno(2.18)$$ 
and the horizon is at 
$$ x^- = -\D = -\lim_{x^+ \to +\infty} \D (x^+)\quad.\eqno(2.19)$$ 
In the limit of a shock-wave matter distribution, for which
$$T_{++}^0 = a \d (x^+ - x_0^+) \quad,\eqno(2.20)$$
we have
$$ M(x^+) =\l a x_0^+ \theta (x^+ - x_0^+)\quad,\quad 
\D(x^+) = {a\over \l^2} \quad.\eqno(2.21)$$
The asymptotically 
flat coordinates $(\h^+,\h^-)$ at the past null infinity $\ci^-$ are given by
$$ \l x^+ = e^{\l \h^+} \quad,\quad x^- = -\D e^{-\l \h^-} \quad,\eqno(2.22)$$
while the asymptotically flat coordinates $(\s^+,\s^-)$ at the future
null infinity $\ci^+$ satisfy
$$ \l x^+ = e^{\l \s^+} \quad,\quad \l (x^- + \D ) = - e^{-\l \s^-}\quad.
\eqno(2.23)$$

Note that a conformal change of coordinates $x^\pm \to \x^\pm$ defines a new 
set of creation and annihilation operators ($b_k^{\dagger} , b_k$) through
$$f_\pm = {1\over\sqrt{2\p}}\int_{0}^{\infty}
{dk\over\sqrt{2\o_k}}\left[ b_{\mp k} e^{-ik\x^\pm} + b_{\mp k}^{\dagger} 
e^{ik\x^\pm}\right]\quad.\eqno(2.24)$$
The old and the new creation
and annihilation operators are related by a Bogoliubov transformation
$$ a_k = S^{-1} b_k S = \int_{-\infty}^{\infty} dq ( b_q \a_{qk} + 
b_q^{\dagger} \b_{qk}^* )\quad,\eqno(2.25)$$
and the new vacuum is given by $ \sket{0_\x} = S \sket{0}$.
Since initially the geometry should be as close as possible to the dilaton 
vacuum,
we take for $\Psi_0$ a coherent state 
$$\Psi_0 = e^A \sket{0_{\h}^+}\otimes \sket{0_{\h}^-}\quad, \eqno(2.26)$$
where $\sket{0_\h} = \sket{0_\h^+}\otimes\sket{0_\h^-}$ is the vacuum for
the ``in" coordinates (2.22), while
$$A = \int_0^{\infty}dk [f_0 (k) a_{-k}^{\dagger} - f_0^{*}(k) a_{-k}] 
\quad,\eqno(2.27)$$ 
and $f_0 (k)$ are the Fourier modes of $f_0 (x^+)$. The choices (2.26) and
(2.27) complete the initial set up of the quantization method of \cite{m96}.

\sect{3. One-loop spectrum}

A one-loop effective metric can be obtained from the 
expression (2.15), which was calculated in \cite{m96}. This gives
$$e^{-\r_0}=e^{-\f_0} = C + b_{\pm}x^\pm - \l^2 x^+ x^- 
- {\k\over 4} \log |\l^2 x^+ x^- | -
 \ha\int_{-\infty}^{x^+} dy^+ (x^+ - y^+)\left({\pa f_0\over \pa
y^+}\right)^2 \quad.\eqno(3.1)$$
The expression (3.1) can be also obtained as a solution 
of the equations of motion of an effective one-loop action
$$S_{eff} = S_0 - {\k\over 4}\int d^2 x \sqrt{-g}R\bo^{-1}R - 
\k \int d^2 x \sqrt{-g}(R \f -(\nabla \f )^2 )\quad, \eqno(3.2)$$
where $S_0$ is the CGHS action (2.1)\cite{bpp}.
By choosing $C=-\frac14 \k [1 - \log (\k/4)] $ one can obtain a
consistent  semiclassical geometry \cite{bpp}. In the case of
the shock-wave matter this geometry is well defined in the $x^+
>0,x^- < 0$ quadrant. For $x^+ < x_0^+$
$$e^{-\r_0}= e^{-\f_0} = C - \l^2 x^+ x^- 
- {\k\over 4} \log (-\l^2 x^+ x^- ) \quad,\eqno(3.3)$$ 
which is defined for $\s \ge \s_{cr}$, where $\s=\log (-\l^2 x^+ x^- )$ is 
the 
static coordinate. At $\s = \s_{cr}$ there is a singularity, and this line is 
interpreted as the boundary of the strong-coupling region. 
 
For $x^+ > x_0^+$ one obtains an evaporating black hole solution
$$e^{-\r_0}=e^{-\f_0} = C + {M\over \l} - \l^2 x^+ (x^- + \D ) 
- {\k\over 4} \log (-\l^2 x^+ x^- ) \quad.\eqno(3.4)$$
The corresponding Hawking radiation flux at $\ci^+$
is determined in the operator formalism by evaluating 
$$ \sbra{\Psi_0} T_{--} (\x^-) \sket{\Psi_0} \quad,\eqno(3.5)$$
where $T_{--}(\x^-)$ is normal ordered with respect to
the asymptotically flat coordinates $\x^{\pm}$  of the metric (3.4) at $\ci^+$.
Since the $\x^{\pm}$ coordinates are the same as the ``out"
coordinates (2.23) for the classical black hole solution, one obtains
by using the point-splitting regularization
$$  \svev{T_{--}(\x^-)} = {\l^2\over 48}\left[ 1 - (1 + 
\l\D e^{\l\s^-})^{-2}\right]\quad.\eqno(3.6)$$
The expression (3.6) gives a late-time flux which corresponds
to a thermal Hawking radiation, with 
the temperature
$T = \l/ 2\p$ \cite{{cghs},{gn}}. The Hawking radiation shrinks the 
apparent horizon of the solution (3.4), so that the apparent horizon line 
meets the curvature singularity in a finite proper time, at 
$$ x_i^+ = {1\over \l^2 \D} \left( -\k/4 + e^{1 + {4\over k}(C + M/\l
 )}\right)\quad,\quad
 x_i^- = {- \D\over 1 - {\k\over 4} e^{-1 - {4\over k}(C + M/\l)}}\quad. 
\eqno(3.7)$$
The curvature 
singularity then becomes naked for $x^+ > x^+_i$. However, a static solution 
(3.3) of the form
$$e^{-\r_0}=e^{-\f_0} = {\hat C}  - \l^2 x^+ (x^- + \D ) 
- {\k\over 4} \log (-\l^2 x^+ (x^- + \D)) \eqno(3.8)$$
can be continuously matched to (3.4) along $x^- = x_{i}^-$
if ${\hat C} =-\frac14 \k [1 - \log (\k/4)]$. A small
negative energy shock-wave emanates from that point, and for $x^- >
x_{i}^-$ the Hawking radiation stops, while the static geometry (3.8) has
a null ADM mass. There is again a critical line $\tilde{\s} =\tilde{\s}_{cr}$,
corresponding to a singularity of the geometry (3.8). Note that the scalar
curvature of (3.8) is bounded at $x^- = x^-_i$, and the singularity comes from
the pathological behaviour of $e^{-\f_0}$, which becomes ill-defined for 
$x^- > x_i^-$. This singularity can be
interpreted as the boundary of the region where higher-order
corrections become important. The spatial geometry of the remnant (3.8)
is that of a semi-infinite throat, extending to the strong coupling region.

Note that the form of the Hawking flux (3.6) is necessary
but not a sufficient 
condition for the Hawking radiation to have 
a Planckian spectrum. In order to determine this, one has to 
evaluate the Bogoliubov coefficients. Since
the ``in" and the ``out" coordinates are exactly the same as in the classical
black hole background case, 
one can conclude that the one-loop Bogoliubov 
coefficients are the same as the classical Bogoliubov 
coefficients calculated in \cite{gn}. 
Consequently, a Planckian one-loop spectrum should be obtained in the 
late-time approximation. However, by a closer examination of the one-loop 
geometry 
one notices that the one-loop horizon is given by the line $x^- = x_i^-$.
This means that
the classical horizon $x^- = -\D$ has undergone
a small shift due to the
one-loop backreaction. The shift $-\D - x_i^-$ is very small, of the
order of $\D e^{-M/\l}$. Still, as we are going to show, this shift will have
a non-trivial consequence for the spectrum of the Hawking radiation.

The Bogoliubov coefficients can be calculated by using the formalism
of \cite{gn}. The ``in" plain-wave basis is given by
$$ u_\o = {1\over\sqrt{2\o}} e^{-i\o\h^-} \quad,\eqno(3.9)$$
while the ``out" basis is given by 
$$ v_\o ={1\over\sqrt{2\o}} e^{-i\o\s^-}\theta (\h_i -\h^-) \quad,
\eqno(3.10)$$
where $\h_i$ is the position of the one-loop horizon, given by 
$$\l \h_i = \log \left[ 1 - {k\over 4}e^{-{k\over 4}(M/\l + C) - 1} \right]
\quad.\eqno(3.11)$$
In the semiclassical approximation, when $M>>\l$,
the shift is an exponentially small quantity. Still, it produces a change in
the classical ``out" basis, for which $\h_i = 0$. 
Following \cite{gn}, one obtains
$$ \a_{\o\o^{\prime}} = -{i\over \p} \int_{-\infty}^{\h_i} d\h^-\, v_\o 
\pa_{\h^-} u_{\o^{\prime}}^{*} \quad,\quad 
\b_{\o\o^{\prime}} = {i\over \p} \int_{-\infty}^{\h_i} d\h^- \, v_\o 
\pa_{\h^-} u_{\o^{\prime}} \quad,\eqno(3.12)$$
so that
$$\a_{\o\o^{\prime}} = {1\over 2\p}\sqrt{\o^{\prime}\over\o}
\int_{-\infty}^{\h_i} d\h^-\,e^{-i\o\s^- + i\o^{\prime}\h^-} \quad,
\eqno(3.13)$$
while 
$$\b_{\o\o^{\prime}}= {1\over 2\p}\sqrt{\o^{\prime}\over\o}
\int_{-\infty}^{\h_i} d\h^-\,e^{-i\o\s^- - i\o^{\prime}\h^-}=
-i\a_{\o -\o^{\prime}}\quad.\eqno(3.14) $$

From (2.23) we obtain
$$ \a^{\pm}_{\o\o^{\prime}} = 
{1\over 2\p}\sqrt{\o^{\prime}\over\o}
\int_{-\infty}^{\h_i} d\h^-\,e^{i{\o\over\l}\log[\l\D (e^{-\l\h^-} -1)] \pm 
i\o^{\prime}\h^-}\quad,\eqno(3.15) $$
where $\a^+ =\a$ and $\a^- = \b$.
The integral in (3.15) can be rewritten as
$$I_1 = \int_0^{x_i} dx (1-x)^a x^{-1-a + b} \eqno(3.16)$$
where $a= i\o/\l$, $b= \pm i\o^{\prime}/\l$, $x_i = e^{\l\h_i}$.
The integral (3.16) is the incomplete Beta function $B(1+a,b-a,x_i)$
\cite{funct}, so that
$$\a^{\pm}_{\o\o^{\prime}} = {1\over 2\p\l}\sqrt{\o^{\prime}\over\o}
(\l\D)^{i\o/\l}
B\left( 1 +i\o/\l, -i(\o \mp \o^{\prime})/\l,x_i\right) \quad.
\eqno(3.17)$$
The incomplete Beta function can be evaluated for $x_i$ close to one via the
expansion \cite{funct}
$$B(1+a,b-a;x)=B(1 +a,b-a)
-{x^{ b-a}\over 1+a}\su_{n=0}^{\infty}{(1+b)_n\over (2+a)_n}(1-x)^{n + 1+ a}
\quad,\eqno(3.18)$$ 
where $(c)_n = {\G ( c + n)\over \G(c)}$. When $x_i =1$, the formula (3.18)
gives the expression for the Bogoliubov coefficients obtained in \cite{gn}.

The number of particles of a given energy emmited at $\ci^+$ is given by
$$ N_\o = \int_0^{\infty} |\b_{\o\o^{\prime}}|^2 d\o^{\prime} \quad,
\eqno(3.19)$$
which is divergent when (3.17) is used. This divergence is the artefact of
the plain-wave basis we are using, 
and can be avoided by using a normalisable basis \cite{hawk}.
A convinient basis is formed by the vawe-packets
$$ v_{j\e} (n) = {1\over{\sqrt \e}}
\int_{j\e}^{(j+1)\e} d\o e^{{2\p in\over\e}\o} v_\o
\quad,\eqno(3.20)$$
which are centered around $\s^- = {2\p n\over \e}$, where $n,j \in {\bf N}$
\cite{gn}.
In the late-time approximmation, i.e. when $n$ is large, the main contribution
in the $\b$-coefficient integral comes from the vicinity of the horizon,
for $\l \h^- = O (\exp (-2\p n/\e))$, so that one can use the
approximation $e\sp{-\l\h^-} - 1 \approx -\l\h^- $, which gives
$$\b_{\o,\o^{\prime}} \approx {1\over 2\p}\sqrt{\o^{\prime}\over\o}
\int_{-\infty}^{\h_i} d\h^-\,e^{i{\o\over\l}\log[-\l^2 \D\h^- ] - 
i\o^{\prime}\h^-}\quad.\eqno(3.21) $$
This is also equivalent to large $\o^{\prime}$ asymptotics of (3.17), since 
the contribution from small $\h^-$ is equivalent to
$\log \o^{\prime}\approx 2\p n/\e$. 
 
The expression (3.21) can be evaluated by using the integral
$$ J_1 = \int_{\e_i}^{\infty} dx x^a e^{-bx} = b^{-1-a} \G (1+a,b\e_i) \quad,
\eqno(3.22)$$
where $\G (\a , x)$ is the incomplete gamma-function, $a=i\o/\l$,
$b=\pm i\o^{\prime}/\l$ and $\e_i = -\l\h_i$. 
The following expansion
$$\G (\a, x) = \G (\a) -\su_{n=0}^{\infty}{(-1)^n x^{\a + n}\over n!
(\a +n)} \quad,\eqno(3.23)$$
can be used to approximate the incomplete gamma function for small
$x$ \cite{funct},
while for large values of $x$ one can use \cite{funct}
$$\G (\a, x) =  x^{\a -1}e^{-x}\left( 1 + {\a - 1\over x} + O(x^{-2})\right)
 \quad.\eqno(3.24)$$
By using the approximation (3.21) and the formula (3.22) we obtain 
$$ \a^{\pm}_{\o\o^{\prime}} \approx 
{1\over 2\p\l} \sqrt{\o^{\prime}\over\o} 
(\l \D)^{i\o/\l}(\pm i\o^{\prime}/\l)^{-1 -i\o /\l} \G (1 + i\o/\l, 
\mp i\o^{\prime}\h_i )\quad. 
\eqno(3.25)$$
Note that the approximation (3.25)
follows from the exact formula (3.17) when $\o^{\prime}$ is large, in 
accordance with the expected late-time asymptotics. 
This can be
seen from (3.18) by approximating each term in (3.18) by its leading large-$b$
expression. In this way one obtains the first two terms of (3.25) when 
expanded via (3.23). 

From (3.25) and (3.23) we obtain
$${\Big |}{\a_{\o\o^{\prime}}\over\b_{\o\o^{\prime}}}{\Big |}^2 \approx
e^{2\o\p/\l}
\left|{\G(1 +i\o/\l) - i{e^{-\p\o/2\l}\over 1 + i\o/\l} (-\o^{\prime}
\h_i )^{1+i\o/\l} + \cdots \over
\G(1 +i\o/\l) + i{e^{\p\o/2\l}\over 1 + i\o/\l} (-\o^{\prime}
\h_i )^{1+i\o/\l} + \cdots }\right|^2
\quad.\eqno(3.26)$$
One can see that 
the expression (3.26) gives in the limit $\h_i \to 0$ a Planckian spectrum 
with the temperature $T=\l/2\p$. However, the $\h_i$ corrections do not
correspond to a Planckian spectrum with a shifted temperature $ T + \d T$,
as one would naively expect. This can be also seen by applying 
the large $\o\sp{\prime}$ formula (3.24) 
to (3.25), so that
$${\Big |}{\a_{\o\o^{\prime}}\over\b_{\o\o^{\prime}}}{\Big |}^2 \approx
\exp \left( - {4\o\over\l\o^{\prime}\h_i}\right)\ne
\exp \left({\o\over T }\right) 
\quad,\eqno(3.27)$$
where $T$ is independent of $\o\sp{\prime}$.
Note that when $\o^{\prime} \to \infty$ the expression (3.27) tends to one,
which in the thermal case would mean that $T\to\infty$.
This indicates that the corresponding Hawking flux will diverge.
We can calculate the Hawking flux by calculating
$N_\o $, since
$$ \svev{T} = \int_0^{\infty}d\o \o N_\o \quad.\eqno(3.28)$$
$N_\o$ can be calculated by going into the
basis (3.20), so that
$$ \b_{\o\o^{\prime}}(n) = {1\over\sqrt\e}\int_{j\e}^{(j+1)\e} d\n 
e^{{2\p in\over\e}\n} \b_{\n\o^{\prime}} \quad,\eqno(3.29)$$
where $\o = j\e$. By applying the expansion (3.23) to $\b_{\n\o^{\prime}}$,
we obtain in the large-$n$ limit
$$\li{ |\b_{\o\o^{\prime}}(n)|^2 =& {1\over \p\l\e}
(e^{2\p\o/\l}- 1)^{-1}
{1\over\o^{\prime}} {\sin^2 \e z/2\over z^2} \cr
+& \left({i\h_i e^{-{\p\o\over 2\l}}\over \p^2\e\o}
{\G (1 + i\o/\l)\over 1 - i\o/\l}e^{i\e(j +\ha)(z - z_1)}
{\sin\ha\e z_1 \sin\ha\e z\over z_1 z } + {\rm c.c.}\right)\cr 
+& O(\h_i^2) \quad,&(3.30)\cr}$$
where c.c. denotes the complex conjugate of the previous term, while
$z = -\frac1\l \log(\o^{\prime}/ \l^2\D) + {2\p n\over\e}$ and
$z_1 = \frac1\l \log(-\h_i \l^2\D) + {2\p n\over\e}$. In order to derive
(3.30) one needs
$$ \int_{j\e}^{(j+1)\e} d\n e^{ia\n} ={2\over a}\sin{a\e\over 2} 
e^{i(j +\ha)\e a} \eqno(3.31)$$
and
$$ |\G (1 + ix)|^2 = {\p x\over \sinh \p x}\quad.\eqno(3.32)$$
The $\o^{\prime}$ integral over the first term in (3.30) gives the
standard thermal contribution.
However, the $\o^{\prime}$ integral over the second 
and higher order in $\h_i$ terms in (3.30)
are all divergent. This is caused by the fact that we have taken only a finite 
number of terms
in the expansion (3.23), which is a good approximation only 
for a sufficiently small $x =\o^{\prime}\e_i$ , or equivalently not too big 
$\o^{\prime}$ ($\o^{\prime}< \e_i^{-1}$). For big 
$\o^{\prime}$ ($\o^{\prime}> \e_i^{-1}$), 
the asymptotic expansion (3.24) is better. Consequently, we
have
$$\li{ N_\o (n) = & \int_{0}^{\e^{-1}} d\o^{\prime}|\b_{\o\o^{\prime}}|^2 +
\int_{\e^{-1}}^{\infty} d\o^{\prime}|\b_{\o\o^{\prime}}|^2 \cr
\approx & \int_{0}^{\e^{-1}} d\o^{\prime}|\b^s_{\o\o^{\prime}}|^2 + 
\int_{\e^{-1}}^{\infty} d\o^{\prime}|\b^l_{\o\o^{\prime}}|^2 &(3.33)\cr}$$
where $\b^s$ and $\b^l$ stand for small and large $\o^{\prime}$ 
approximations, respectively. 
The first integral in (3.33) has the integrand given by
(3.30), and because of the cut-off in $\o^{\prime}$, it is finite. 
In the second integral
we use the approximation (3.24), so that  
$$ \b^l_{\o\o^{\prime}} = {1\over 2\p}\sqrt{\o^{\prime}\over\o}  
(\l\D)^{i\o/\l} {x_i^{-i\o^{\prime}/\l}\over -i\o^{\prime}}
\exp( i(\o/\l)\log \e_i)\quad.\eqno(3.34)$$
This gives a logarithmic divergent contribution. However, when $\e_i =0$,
the large $\o^{\prime}$ asymptotics changes by a phase-factor 
$\exp[-i(\o/\l)\log(\o^{\prime}/\l)]$, which gives a finite $N_\o (n)$.

Note that the one-loop flux is also given by 
the expectation value
(3.6), which is finite. From the Bogoliubov coefficients
analysis it follows that a finite flux is possible only if the shift
$\e_i \le 0$, while the one-loop geometry seems to suggest that $\e_i > 0$.
The way out of this paradox is provided by the fact that 
the line $x^- = x^-_i$ gets very close to the strong-coupling region. 
Therefore
the one-loop geometry is not a good approximation there, and hence
our assumption that the effective horizon was given by the line
$x^- = x^-_i$ was not a good assumption. 

\sect{4. Two-loop spectrum}

A two-loop metric can be obtained by truncating the expansion (2.13) after
$n=2$. Outside of the region occupied by the matter pulse centered around
$ x_0^+ $,
the metric can be written as \cite{mr}
$$ds^2 = - e^{\f_2}dx^+dx^-\quad;$$
$$ e^{\f_2} = e^{\f_0} \left[ 1 + e^{2\f_0}(C_- + C_+ (x^+ - x_0^+)^2
\theta (x^+ - x_0^+)) \right]\quad, \eqno(4.1)$$
where $e^{\f_0}$ is the one-loop dilaton solution
$$e^{-\f_0} = C - a (x^+ - x_0^+) \theta (x^+ - x_0^+) - \l^2 x^+ x^- 
-\frac{\k}4 \log |\l^2 x^+ x^-| \quad,\eqno(4.2)$$
whith $a = \l\sp 2 \D$ and
$C_\pm$ are integration constants, analogous to the constant $C$ of the
one-loop solution. The difference now is that $C_+$ depends on the matter 
pulse profile, and it is negative for very narrow pulses (those
which are shorter than
a critical length defined by the momentum cut-off \cite{mr}) and positive
otherwise. The constant $C_-$ is matter independent, and its value is
regularization dependent. Consistent semiclassical geometries appear
for $C_+ \ge 0$ \cite{mr}.

As in the one-loop case, the relevant quadrant is
$x^+ \ge 0,x^- \le 0$. For $x^+ < x_0^+$ the solution (4.1) is static,
and it can be written as
$$ e^{\f_2} = e^{\f_0} \left[ 1 - e^{2\f_0}\a^2 \right]\quad,\quad
e^{-\f_0} = C + e^\s -\frac{\k}4 \s \quad,\eqno(4.3)$$
where $C_- = -\a^2 $ and
$\s =\log (-\l^2 x^+ x^-)$ is the static coordinate. The solution
(4.3) describes
a two-loop corrected dilaton vacuum. The corresponding scalar curvature 
diverges at the line
$$ e^{-\f_0} - \a = 0 \quad,\eqno(4.4)$$
which
corresponds to the one-loop singularity line with $C$ replaced by 
$C - \a$. The semiclassical geometry will be defined
for $\s \ge \s_c$, where $C + e^{\s_c} -\frac{\k}4 \s_c = \a$. The curvature
singularity will be absent for $C\ge \a -\frac{k}4 ( 1- \log\frac{k}4)$.
Therefore if we want to avoid a naked singularity at two loops, the
one-loop $C$ has to be increased to $C= \a -\frac{k}4( 1
- \log\frac{k}4)$. Note that in the case when $C_- = \a\sp 2$, there is no
curvature singularity, and the only singularity comes from the one-loop
critical line determined by the second formula in (4.3).

For $x^+ > x_0^+$ the solution (4.1) becomes 
$$ e^{\f_2} = e^{\f_0} \left[ 1 - e^{2\f_0}(\a^2 - C_+ (x^+ - x_0^+)^2)
 \right]\quad, \eqno(4.5)$$
where 
$$e^{-\f_0} = C + {M\over\l} - \l^2 x^+ (x^- + \D) 
-\frac{\k}4 \log(-\l^2 x^+ x^-)\quad.\eqno(4.6)$$
It describes a two-loop corrected evaporating
black hole geometry. The curvature singularity line is given by
$$ e^{-2\f_0} - \a^2 + C_+ (x^+ - x_0^+)^2 = 0 \quad.\eqno(4.7)$$
When $C_+ = 0$, the equation (4.7) becomes a shock wave 
singularity equation (4.4), and it describes
a singularity line of the one-loop metric with a smaller ADM mass
$ C + {M\over \l} - \a$. The apparent horizon line is given by the equation
$\pa_+ \f_2 = 0$, which can be rewritten as
$$  x^- + \D +{k\over 4\l^2 x^+} = {-2C_+ (x^+ - x_0^+)e^{-\f_0}/\l^2\over
e^{-2\f_0} - 3\a^2 + 3C_+ (x^+ - x_0^+)^2 } \quad.\eqno(4.8)$$
In the shock-wave limit ($C_+ =0$) the intersection point of the apparent 
horizon and the curvature singularity line 
is given by the one-loop expressions (3.7). 
The change in the intersection point when $C_+ \ne 0$ can be 
evaluated perturbatively \cite{mr}, and it is given by
$$\li{ \d x_i^- =& -{C_+\over 2\a} {(x_i^+ - x_0^+)^2\over ax_i^+} x_i^- 
+ O(\b^4)\cr
 \d x_i^+ =& -{2\l^2 C_+\over \k\a} {(x_i^+ - x_0^+)^2\over a} x_i^-
x_i^+ + O(\b^4) \quad.&(4.9)\cr}$$
The line $x^- = x_h =x_i^- + \d x_i^-$ can be considered as the horizon 
of the
two-loop semiclassical geometry. Clearly the quantum corrections have shifted
the position of the horizon, which will affect the Hawking radiation. In
the region $x^- > x_h$ a naked singularity will appear, unless we
impose an
appropriate boundary condition. In the shock-wave case, one can
impose a static solution (4.3) for $x^- > x_h$ 
$$e^{\f_2} = e^{\f_0} \left[ 1 - e^{2\f_0}\a^2 \right]\quad,\quad
e^{-\f_0} = {\hat C}  - \l^2 x^+ (x^- + \D) 
-\frac{\k}4 \log(-\l^2 x^+ (x^- + \D))\quad,\eqno(4.10)$$
which can be continuously matched to (4.5) at $x^- =x_h$ if 
$${\hat C}= -{\k\over 4}(1-\log{\k\over 4}) - \a \quad.\eqno(4.11)$$
However, in contrast to the one-loop case, the metric (4.10) has a curvature
singularity at $\s = \s_{cr}$, and the naked singularity is not removed. 
When $C_+ < 0$, there is no static 
two-loop dilaton vacuum solution for $x^- > x_h$
which can be continuously matched to (4.3)
and the naked singularity remains. However, when $C_+ > 0$, the naked 
singularity lays in the strong-coupling region, and therefore it can be
ignored since it lays outside of the region of validity of the metric (4.1).
Also note that in the case when $C_- >0$, the solution with $C_+ >0$ does not 
have a curvature singularity.
As a consequence of these properties the Hawking flux will be
free of singularities only for $C_+ > 0$. 
  
The asymptotically flat coordinates $(\s^+,\s^-)$ at $\ci^+_R$ are
given by
$$ \l x^+ = e^{\l\s^+} \quad,\quad \l(x^- + \D ) = - e^{-\l \tilde{\s}}
\quad,\eqno(4.12)$$
where
$$ \tilde{\s} = \s^- - {C_+\over 2\l^3} e^{2\l\tilde{\s}}\quad. \eqno(4.13)$$
This can be written as
$$\s^- = -{1\over \l} \log [-\l(\D + x^-)] +
{C_+\over 2\l^5} (\D + x^-)^{-2} \quad,\eqno(4.14)$$
so that a non-zero correction appears at two loops. 
As a direct consequence, 
the Hawking flux at $\ci^+$ will not have a zero-loop form (3.6), which can 
be seen by evaluating  
$$ \svev{T_{--}}|_{\ci^+} = -{1\over 48}\left(
{\h^{\prime\prime\prime} \over\h^{\prime}} - 
\frac32 \left({\h^{\prime\prime}\over\h^{\prime}}\right)^2\right) 
\quad,\eqno(4.15) $$
where
the primes denote $d/d\s^-$. One then obtains \cite{mr}
$$ \ct =\svev{T_{--}}|_{\ci^+} = {\l^2\over 24}
{y^4[y^4 (-\D y + \ha\D^2 ) - C_+ P_4 (y) + C_+^2 P_2 (y) ]\over
(y - \D )^2 ( C_+ + y^2 )^4 }\quad, \eqno(4.16)$$
where $y =x^- + \D$, $P_2 = -2y^2 + 3\D y -3/2 \D^2$ and
$P_4 = y^2 ( -4y^2 + 10\D y -5 \D^2)$. From (4.16) one can see that
$\ct (y)$ does not diverge for $C_+ \ge 0$. In that case $\ct (y)$ goes to
zero for late times ($y \to 0$), indicating that higher-order loop corrections
can turn off the Hawking radiation. This also means that the spectrum
of the Hawking radiation can not be thermal for very late times.

Relation (4.14) clearly shows that the two-loop ``out" coordinates are not
the same as the analogous zero and one loop coordinates, which indicates 
that the spectrum of the radiation will change. This can be confirmed
by evaluating the Bogoliubov coefficients in the late-time approximation.
The two-loop Bogoliubov coefficients can be evaluated from the formulas
(3.13-14), where now $\l\s^- =-\log[\l\D (e^{-\l\h^-} -1)]+
{C_+\over 2\l^4 \D^2} (e^{-\l\h^-} -1 )^{-2} $ and
$$ \l\h_i = \log \left[ 1 - {k\over 4}e^{-{k\over4}(M/\l + C) - 1} 
\right]
+ C_+ {(x_i^+ - x_0^+ )^2\over 2\a a x_i^+}\quad,\eqno(4.17)$$
so that
$$ \a^{\pm}_{\o^{\prime}\o} = {1\over 2\p}\sqrt{\o^{\prime}\over\o}
\int_{-\infty}^{\h_i} d\h^-\,
\exp{\Big(} i{\o\over\l}\log[\l\D (e^{-\l\h^-} -1)] 
- {i\o C_+\over 2\l^5 \D^2} (e^{-\l\h^-} -1 )^{-2} \pm 
i\o^{\prime}\h^- {\Big )}\,.\eqno(4.18) $$

For late times, we can use the approximation 
$e\sp{-\l\h^-} - 1 \approx -\l\h^- $,
so that
$$\a^{\pm}_{\o\o^{\prime}} \approx {1\over 2\p}\sqrt{\o^{\prime}\over\o}
\int_{-\infty}^{\h_i} d\h^-\,
\exp\left( i{\o\over\l}\log (-\l^2 \D \h^- ) -
{i\o C_+\over 2\l^5 \D^2} (\l\h^- )^{-2} \pm
i\o^{\prime}\h^- \right) \quad.\eqno(4.19) $$
Equation (4.19) can be rewritten as
$$\li{ \a^{\pm}_{\o\o^{\prime}} \approx & {1\over 2\p\l}
\sqrt{\o^{\prime}\over\o}(\l\D)^{i\o/\l}
\int_{\e_i}^{\infty} dy\, y^a e^{-by - c y^{-2}}\cr
\approx & {1\over 2\p\l}\sqrt{\o^{\prime}\over\o}(\l\D)^{i\o/\l} J_2 (a,b,c)
 \quad,&(4.20)\cr} $$
where $a=i\o/\l$, $b=\pm i\o^{\prime}/\l$, $\e_i = -\l\h_i $ and
$c = i\o C_+ / 2\l^5 \D^2$. Since we are interested 
in the large-$b$ asymptotics,
we can analyze $J_2$ by using the method of stationary points. We can rewrite
$J_2$ as
$$J_2 = \int_{\e_i}^{\infty} dy\, y^a e^{-f(y)}\quad,\eqno(4.21)$$
where $f(y)= by + cy^{-2}$. $f(y)$ has a stationary point at 
$y = (2c/b)^{\frac13}$, which for a sufficiently large $b$ will lay outside of 
the interval $[\e_i,\infty)$. Hence the main contribution to 
$J_2$ will come from the vicinity of the point $y=\e_i$, so that
as $b\to \infty$
$$J_2 \approx {1\over b}(\e_i)^a \exp( -b\e_i - c \e_i^{-2})
\quad.\eqno(4.22) $$
Note that the argument we have used to derive (4.22) clearly holds for $a,b,c$
real and positive. When $a,b,c$ are complex, an analogous argument applies
(see \cite{olver} and the appendix), 
and (4.22) is still valid. We also show in the appendix that
the same method can be applied to the
exact expression (4.18), and the result (4.22) is again obtained when $b$ is
large.

The expression (4.22) has the same asymptotics as the incomplete gamma 
function (3.22), and
therefore we will encounter the same problem as in the one-loop case.
Since $N_\o (n)$ will then diverge, the corresponding Hawking flux 
will also diverge for late times. On the other hand, the Hawking flux 
from the expectation value (4.16)
is finite for late times, and we have the same discrepancy as in the one-loop
case. Since the line $x\sp - = x_i\sp - + \d x_i\sp - $, 
which is suggested by the semiclassical geometry
as the horizon, gets close to the border of the strong-coupling 
region where the spacetime geometry is not well-defined, 
we can invoke the same argument as in the one-loop case, and resolve
the discrepancy by taking $\e_i \le 0$. In that case the
stationary-point method gives thermal Bogoliubov coefficients 
(see the appendix),
and we obtain a Planckian late-time spectrum 
in the leading-order approximation with the temperature 
$T = 3\l/ 2\p$. This increase in the temperature is an expected 
back-reaction effect. However, an unexpected feature is that the increase
is not perturbative in $C_+$. A related problem is that the Hawking flux 
(4.16) deviates only slightly for small enough $C_+$
from the zero-loop flux which corresponds to the temperature $T = \l/2\p$.
Note that for big enough
$C_+$ there is a late time $y_0$ for which $\ct (y_0)$ coincides 
with the thermal flux corresponding to the temperature $T = 3\l/2\p$. 
In this case one can argue that there is an agreement with the
Bogoliubov coefficients calculation when $y\approx y_0$, 
while for the times $y > y_0$, the
discrepancy can be justified by the argument that the background
geometry is not well-defined (strong-coupling region)
and hence the Bogoliubov coefficients
are not well defined there. Still, there is a discrepancy in the weak-coupling
region for small $C_+$, and a possible explanation is that the approximation
(A.15) is not good enough, and that the $a_2$ and higher-order terms should 
have been taken into account.

\sect{5. Conclusions}

The results of our analysis indicate that small back-reaction corrections
in the background geometry can induce large changes in the Hawking radiation
spectrum. 
We have seen that a small positive shift in the position
of the classical horizon gives a non-Planckian late-time Hawking radiation 
spectrum whose flux diverges. On the other hand, a small negative or a null 
shift gives a Planckian late-time spectrum. Consistency considerations
then select a null or a negative shift as an effective horizon shift, contrary
to what is suggested by the semiclassical geometry. This is
explained  by the breakdown of
the semiclassical approximation for late times near the classical horizon.

In the one-loop case the Bogoliubov coefficients give a
late-time Planckian spectrum which is consistent with the operator
quantization Hawking flux only for times $x\sp - < x_i\sp -$. After that the 
BPP solution gives a zero flux, which indicates that one enters a non-thermal
regime. It is plausibile that the higher-order stationary point contributions
to the Bogoliubov coefficients may account for this.
In the two-loop case 
the relation between the late-time spectrum and the 
operator quantization
flux is similar but slightly more complex. The consistency again requires
that the time is not too late, and the new requirement is that the constant 
$C_+$ is not too small. It is likely that this restriction
is caused by the low-order approximation we are using to calculate the 
two-loop Bogoliubov coefficients. Including the higher-order terms in the 
stationary-point
approximation may also account for a non-Planckian nature of the very late
time spectrum, which is suggested by the behaviour of the operator
quantization Hawking flux $\ct$. However, a further investigation is
necessary in order to clarify these issues.

Note that when $C_+ < 0$, there is a total disagreement between the flux
from the Bogoliubov coefficients and the operator quantization flux $\ct$. 
In that case
$\ct$ diverges to minus infinity for late times. This cannot be reconciled
neither with the positive horizon shift case (the corresponding 
flux diverges to plus infinity), nor with the negative horizon shift case 
(the corresponding flux is finite and positive).
However, $C_+ < 0$ solution can be considered as an unphysical solution,
since it requires a matter pulse whose width is shorter than the 2d analog
of the Planck length \cite{mr}.

\sect{Acknowledgements}

\noindent We would like to thank MNTRS for a financial support and 
Olle Brander for useful discussions.

\sect{Appendix}

In this appendix we give some relevant theorems from \cite{olver}, and
use them to obtain the asymptotic expressions for the relevant integrals.

In the one-loop case, the relevant integral is
$$ I_1 (b) = \int_0^{x_i} d\t\, \t^{b-a-1} (1-\t)^a \eqno(A.1)$$
where $x_i =\exp(-\e_i )$, $a =i\o/\l$, $b = \pm i\o^{\prime}/\l$. 
This integral can be rewritten as
$$I_1 (b) =x_i^{b-a} \int_0^{\infty} dt e^{-bt} (e^t - x_i )^a \eqno(A.2)$$
where $\t= x_i e^{-t}$. Now we use a theorem from \cite{olver}, page 108

Theorem (1):  Let $ I(z) = \int_0^{\infty} dt e^{-zt} q(t)$, 
such that $q(t)$ is
real or complex, $|q^{(s)}(t)| \le A_s e^{\s|t|}$ ($\s \in {\bf R}$) and
$q(t)$ holomorphic in the sector $S=\{\a_1 \le Arg\, t \le \a_2\}$ 
($\a_1 <0$, $\a_2 > 0$). Then as $z \to \infty$
$$ I(z) \approx \su_{s=0}^{\infty} {q^{(s)}(0)\over z^{s+1}}\quad, 
\eqno(A.3)$$
in the sector $S_\p =\{-\a_1 - \p/2 \le Arg\, z \le -\a_2 + \p/2\}$.

In the case of the integral (A.2), $q(t) = (e^t - x_i )^a$, and the conditions
of the theorem (1) are satisfied, so that as $b \to \infty$
$$ I_1 (b) \approx x_i^{b-a}{(1-x_i )^a\over b} + O(b^{-2}) \quad.\eqno(A.4)$$
Note that the expression (A.4) has the same large-$b$ asymptotics as (3.22)
for $x_i$ close to one.

When $x_i =1$, then a generalization of the theorem (1) applies \cite{olver}, 
page 114

Theorem (2): Let $q(t)$ be holomorphic in $S$ and $q(t) = O(e^{\s|t|})$ as 
$t\to\infty$. Let
$$q(t) \approx \su_{s=0}^{\infty} a_s t^{(s+\n -\m)/\m} $$
for $t\to 0^+$ and $t\in S$, where $\m >0$ and ${\rm Re}\,\n >0$. Then
$$ I(z) \approx \su_{s=0}^{\infty}\G \left( {s+\n \over\m}\right)
{a_s\over z^{(s+\n)/\m}} \eqno(A.5)$$ 
as $z\to \infty$ in the sector $S_\p$.

In our case, $q(t)=(e^t - 1)^a \approx \su_{s=0}^{\infty} a_s t^{s+a} $,
so that $\m =1$ and $\n = 1 + a= 1+ i\o/\l$, and hence we can
use the theorem (2). Therefore, as $b \to \infty$
$$ I_1 (b) \approx {\G (1+a)\over b^{1+a}}\left( 1 + O(b^{-1}) \right) 
\quad,\eqno(A.6)$$
whose leading-order term is (3.22) for $\e_i =0$.

In the two-loop case, we have
$$\li{ I_2 (b) =& \int_0^{x_i} d\t\, \t^{b-a-1} (1-\t)^a 
\exp \left[ -c(\t^{-1} -1)^{-2}
\right]\cr 
 =& x_i^{b-a} \int_0^{\infty} dt e^{-bt} (e^t - x_i )^a 
\exp \left[ -c(x_i^{-1}e^{t} -1)^{-2}\right]
\quad.&(A.7)\cr}$$
When $x_i <1$, then
$$q(t)=(e^t - x_i )^a \exp \left[ -c(x_i^{-1}e^{t} -1)^{-2}\right]
 \approx \su_{s=0}^{\infty} a_s t^{s} $$
so that the theorem (1) applies, and hence
$$ I_2 (b) \approx x_i^{b-a} {(1-x_i )^a\over b}
\exp\left[-c\left({x_i\over 1-x_i}\right)^2 \right] + O(b^{-2}) \quad.
\eqno(A.8)$$
It is easy to see that the large-$b$ asymptotics of the expressions (A.8) 
and (4.22) coincide for $x_i$ close to one.

When $x_i =1$, then $q(t) \approx t^a \exp(-ct^{-2})$ as $t \to 0^+$, so that
neither theorem (1) nor theorem (2) can be used. One can then use a
theorem from \cite{olver}, pg 127

Theorem (3): Let
$$ I(z) = \int_a^{b} dt\, e^{-zp(t)} q(t)
 = \int_{C_{ab}} dt\, e^{-zp(t)} q(t)$$
such that

(i) $p(t)$ and $q(t)$ are holomorphic and single-valued in a domain $T$

(ii) curve $C_{ab} \in T$ 

(iii) $p^{\prime} (t)$ has a simple zero $t_0 \in C_{ab}$

(iv) $\theta = Arg\, z \in [\theta_1 ,\theta_2 ]$, $|z| \ge Z$, 
$\theta_2 - \theta_1 < \p$, $Z > 0$, $I(z)$ converges absolutely and
uniformly at $a$ and $b$.

(v) Re$\{ e^{i\theta}(p(t) - p(t_0))\}$ is positive on $C_{ab}$, except
at $t_0$, and is bounded away from zero uniformly with respect to
$\theta$ as $t\to a$ or $b$ along $C_{ab}$.

\noindent Then as $z\to \infty $
$$ I(z) \approx 2e^{-zp(t)}\su_{s=0}^{\infty} \G (s + 1/2)
{a_{2s}\over z^{s+1/2}} \quad,\eqno(A.9)$$
for $\theta_1 \le Arg\, z \le \theta_2$, where
$$ a_0 = {q\over\sqrt{2p^{\prime\prime}}} \quad, \quad 
a_2 = \left[ 2q^{\prime\prime} - {2p^{\prime\prime\prime}q^{\prime}\over
p^{\prime\prime}} + \left( {5p^{\prime\prime\prime 2}\over
6p^{\prime\prime 2}} - {p^{iv}\over 2p^{\prime}}\right)q \right]
(2p^{\prime\prime})^{-3/2}\quad,\eqno(A.10) $$
and so on.

In the case of the integral (A.7) when $x_i =1$ one can take 
$$p(t) =t + {c\over b}(e^t - 1)^{-2}  \quad,\quad q(t) = (e^t - 1)^a \quad.
\eqno(A.11)$$
However, theorem (3) cannot be applied directly, since the role of $z$ is
played by $b$, and in the choice (A.11) $p(t)$ depends on $b$.
In order to obtain a proper $p(t)$, note that if we rewrite $I_2$ as
$$ I_2 (b) = \int_0^{\infty} dt \, e^{-f(t)} \eqno(A.12)$$
where $f(t) = -a \log (e^t -1) + bt + c(e^t - 1)^{-2}$, then a main
contribution to $I_2$ comes from the stationary point $f^{\prime} (t_0) =0$,
which for large $b$ is given as $t_0 \approx (2c/b)^{1/3}$.
Since then $t_0$ is close to zero, one can approximate $f(t)$ with its
small $t$ asymptotics, which is the late-time approximation we used in
sections 3 and 4. Therefore as $b\to \infty$ 
$$bp(t) \approx bt + ct^{-2}  \quad,\quad q(t) \approx t^a \quad,
\eqno(A.13)$$
and we make a rescaling $t = \t (c/b)^{\frac13}$, so that
$$ bp(t) \approx z \tilde{p} (\t) = z ( \t + \t^{-2} )\quad,$$ 
where $z = (b^2 c)^{1/3}$. Consequently
$$ I_2 \approx (c/b)^{(a + 1)/3} \int_0^{\infty} d\t \, 
e^{-z(\t + \t^{-2})} \t^a \quad,\eqno(A.14)$$
and the theorem (3) can be now applied. We then obtain in the leading-order
approximation ($a_0$ term)
$$\left| {\a_{\o\o^{\prime}}\over \b_{\o\o^{\prime}}}\right|^2 \approx
|(-1)^{a/3}|^2 =\exp(2\o\p/3\l) \quad,\eqno(A.15)$$
so that the leading-order
two-loop Hawking temperature is given by $T=3\l/2\p$.

\end{document}